%% file: main.tex
\documentclass[sigplan]{acmart}
\AtBeginDocument{%
  \providecommand\BibTeX{{%
    \normalfont B\kern-0.5em{\scshape i\kern-0.25em b}\kern-0.8em\TeX}}}

\usepackage[utf8]{inputenc}
\usepackage{natbib}
\usepackage{amsmath,amsfonts}
\usepackage{algorithmicx}
\usepackage{graphicx}
\usepackage{textcomp}
\usepackage{xcolor}

\usepackage{hyperref}
\usepackage{glossaries}
\usepackage{graphicx}
\usepackage{listings}
\usepackage{csquotes}
\usepackage{tabularx}
\usepackage[tableposition=above]{caption} 
\usepackage{multirow}
\usepackage{enumitem}

\usepackage[export]{adjustbox}

\usepackage{subcaption}

\usepackage{lipsum}

\input{src/inc/Listings.tex}

\input{src/inc/Commands.tex}

\input{src/inc/Glossaries.tex}

\AtBeginDocument{%

}

\setcopyright{none}
\renewcommand\footnotetextcopyrightpermission[1]{}

\begin{document}

\emergencystretch 3em

\title{Building Whitespace-Sensitive Languages Using Whitespace-Insensitive Components}

\author{Alexander Hellwig}
\orcid{0009-0001-1698-4054}
\affiliation{%
  \institution{Software Engineering, RWTH Aachen University}
  \city{Aachen}
  \country{Germany}
}

\author{Nico Jansen}
\orcid{0000-0001-5199-8323}
\affiliation{%
  \institution{Software Engineering, RWTH Aachen University}
  \city{Aachen}
  \country{Germany}
}

\author{Bernhard Rumpe}
\orcid{0000-0002-2147-1966} 
\affiliation{%
  \institution{Software Engineering, RWTH Aachen University}
  \city{Aachen}
  \country{Germany}
}

\renewcommand{\shortauthors}{Alexander Hellwig, Nico Jansen, and Bernhard Rumpe}

\input{src/pre/00.abstract}


\maketitle

\pagestyle{plain}

\input{src/tex/01.introduction}
\input{src/tex/02.background}
\input{src/tex/03.approach}
\input{src/tex/04.evaluation}
\input{src/tex/06.discussion}
\input{src/tex/05.relatedWork}
\input{src/tex/07.conclusion}

\begin{acks}
Funded by the Deutsche Forschungsgemeinschaft (DFG, German Research Foundation) - 459671966 
\end{acks}

\balance

\bibliographystyle{ACM-Reference-Format}
\bibliography{main}

\end{document}

%% file: src/inc/Listings.tex
\usepackage{listings}
\newenvironment{ownfigure}[0]%
{\begin{figure}[htb!]}
{\end{figure}}

\usepackage{color}
\definecolor{DarkRed}{rgb}{0.75,0,0}
\definecolor{Lightgreen}{rgb}{0.588,1.0,0.588}
\definecolor{DarkGreen}{rgb}{0,0.5,0}
    
\lstdefinelanguage{MontiArc}[]{Java}{
  morekeywords={component, port, in, out, inv, package, import, connect, autoconnect}
}

\lstdefinelanguage{myJava}[]{Java}{
  commentstyle=\color{DarkGreen}\itshape 
}

\lstdefinelanguage{MontiArcAutomaton}[]{Java}{
  morekeywords={component, port, in, out, inv, package, import, connect,
  autoconnect, automaton, state, ocl, java, initial, final,
  noCompletion, chaosCompletion, var, mode, activate, transitions,
  modetransitions}, commentstyle=\color{DarkGreen}\itshape }

\lstdefinelanguage{MCConfig} { 
    morekeywords={config, Require, Model} 
}

\lstdefinelanguage{Manifest} { 
    morekeywords={Manifest, Bundle, ManifestVersion, Name, SymbolicName,
      Version, Require
    } 
}

\lstdefinelanguage{mcGrammar}[]{}{
  morekeywords={
    grammar, package, path, parser, lexer, nows, noslcomments, nomlcomments, 
    noident, nostring, noanything, nocharvocabulary, dotident, identrule,
    xmlcomments, hashcomments, texcomments, freemarkercomments, concept, 
    globalnaming, define, usage, options, true, false, protected, ident, 
    compilationunit, symbol, implements
  }
}

\lstdefinelanguage{mcLng}[]{}{
  morekeywords={
    dsltool, language, package, path, parser, root, parsingworkflow, 
    rootfactory, lexer, nows, noslcomments, nomlcomments, noident, nostring,
    dotident, concept, globalnaming, define, usage, options, true, false, 
    protected, ident
  }
}

\lstdefinelanguage{mcManifest}[]{}{
  morekeywords={
    bundle, Bundle, Name, SymbolicName, true, false, Main, Class, 
    Version, Activator, Localization, Require, 
    Exclude, Eclipse, LazyStart, Vendor, Export, Package, 
    ClassPath
  }
}

\lstdefinelanguage{Alloy}[]{Java}{
commentstyle=\color{DarkGreen}\itshape,
  morekeywords={abstract,sig,->,fact,pred,fun,run,for,iff,
  not,no,one,all,some,lone,\#,set,in,and,or,but,exactly,none,univ,Int,assert,check},
  otherkeywords = {[2]????},
    morekeywords = {[2]????},
    keywordstyle={[2]\color{blue}},
    otherkeywords = {[3]????,<,<->,->, &, |, =, !=, !,<:,~},
    morekeywords = {[3]????,<,<->,->, &, |, =, !=, !,<:,~},
    keywordstyle={[3]\color{blue}}
  }

\lstdefinelanguage{mccd}[]{Java}{
  morekeywords={classdiagram,abstract,<<singleton>>,class,int,String,
  association,composition,extends}
}

\lstdefinelanguage{FreeMarker}[]{}{
  keywordsprefix={\#},
  keywords={in},
  commentstyle=\color{DarkGreen}\itshape }

\lstdefinelanguage{Mona}[]{}{
  morekeywords={ex0,all0,ex1,all1,ex2,all2,var0,var1,var2,pred,in,notin,include,union,inter,empty,assert},
  morecomment=[l]{\#},
  commentstyle=\color{DarkGreen}\itshape,
  otherkeywords = {[2]????,next,boolean,init,case,esac},
  morekeywords = {[2]????,next,boolean,init,case,esac},
  otherkeywords = {[3]????,<,<=>,=>, &, |, =, !=, !},
  morekeywords = {[3]????,<,<=>,=>, &, |, =, !=, !},
}

\lstdefinelanguage{myPython}[]{Python}{
  morekeywords={assert},
  morecomment=[l]{\#},
  commentstyle=\color{DarkGreen}\itshape,
}

\lstdefinelanguage{GeneratorConfiguration}[]{Java} {
  morekeywords={
    template, 
    generator, 
    ast, 
    runtime},
}

\lstdefinelanguage{ApplicationConfiguration}[]{Java} {
  morekeywords={
    application,
    behaviors,
    bindings,
    classdiagrams,
    components,
    factories,
    generators,
    map,
    to},
}

\lstdefinelanguage{Isabelle}[]{} {
    morekeywords={
        datatype,
        typedef},
}

\lstdefinelanguage{Task}[]{}{
  morecomment=[l][\color{gray}]{/}{/},
  morekeywords={
    Task,
	Find,
	Place,
	Fill,
	Set,
	Guided,
	Wait,
	Go,
	into,
    can,
    cook,
    with,
    onto,
    to,
    full,
    Drive,
    import
  }
}

\lstdefinelanguage{Context}[]{}{
  morecomment=[l][\color{gray}]{/}{/},
  morekeywords={
    Machines,
    Storages,
    Utensils,
    Ingredients,
    top,
    bottom,
    right,
    left,
    front,
    back,
    middle,
    can,
    cook,
    fry,
    Tools,
    Fasteners,
    WardrobeParts,
    above,
    on,
    of,
    import,
    in,
    is,
    leftOf
  }
}

\lstdefinelanguage{GuiDSL}[]{}{
  morekeywords={
	webpage,
	button,
	container,
	head,
	row,
	helpbutton,
	body,
	datatable,
	column,
	card,
	columns,
	textoutput,
	pie,
    all,
    foreach,
    in,
    label
  }
}

\lstdefinelanguage{GUIDSL1}[]{}{
  morekeywords={
	webpage,
	button,
	container,
	head,
	row,
	helpbutton,
	body,
	datatable,
	column,
	card,
	columns,
	textoutput,
	pie,
    all,
    foreach,
    in,
    label
  }
}

\lstdefinelanguage{GUIDSL2}[]{}{
  morekeywords={
	page,
    component,
    iterate,
    guard,
    else,
    package,
    import
  }
}

\lstdefinelanguage{TypeScript}[]{}{
  morekeywords={
	import,
	from,
	template,
	selector,
	imports,
	export,
	class
  }
}

\usepackage{tikz}

%% file: src/inc/Commands.tex
\usepackage{xspace}
\usepackage{ifdraft}

\usepackage{prettyref}
\newrefformat{chap}{Chapter~\ref{#1}}
\newrefformat{sec}{Section~\ref{#1}}
\newrefformat{subsec}{Section~\ref{#1}}
\newrefformat{eq}{Equation~(\ref{#1})}
\newrefformat{fig}{Figure~\ref{#1}}
\newrefformat{ex}{Example~\ref{#1}}
\newrefformat{list}{List~\ref{#1}}
\newrefformat{tab}{Table~\ref{#1}}
\newrefformat{enum}{Case~(\ref{#1}.)}
\newrefformat{def}{Definition~\ref{#1}}
\newrefformat{le}{Lemma~\ref{#1}}
\newrefformat{th}{Theorem~\ref{#1}}

\makeatletter
\newcommand*{\etc}{%
  \@ifnextchar{.}%
  {\textit{etc}}%
  {\textit{etc.}\@\xspace}%
}
\makeatother

\definecolor{se-green}{RGB}{0,128,0}
\definecolor{se-blue} {RGB}{0,0,204}

\newcounter{requirement}[section]

%% file: src/inc/Glossaries.tex
\newacronym{BPMN}{BPMN}{Business Process Model and Notation}

\newacronym{BPD}{BPD}{Business Process Diagram}

\newacronym{CTL}{CTL}{Computation Tree Logic}

\newacronym{EBNF}{EBNF}{Extended Backus-Naur Form}
\def\ebnf{\gls{EBNF}\xspace}

\newacronym{MDSE}{MDSE}{Model-Driven Software Engineering}

\newacronym{MBSE}{MBSE}{Model-Based Software Engineering}

\newacronym{OCL}{OCL}{Object Constraint Language}

\newacronym{UML}{UML}{Unified Modeling Language}

\glsunset{UML}

\newacronym{UML/P}{UML/P}{\glspl{UML}/P}

\newacronym{CD}{CD}{Class Diagram}

\newacronym{CD4A}{CD4A}{Class Diagram for Analysis}

\newacronym{OD}{OD}{Object Diagram}

\newacronym{OM}{OM}{Object Model}

\newacronym{AST}{AST}{Abstract Syntax Tree}
\def\ast{\gls{AST}\xspace}

\newacronym{DSL}{DSL}{Domain-Specific Language}
\def\dsl{\gls{DSL}\xspace}
\def\dsls{\glspl{DSL}\xspace}

\newacronym{GPL}{GPL}{General Purpose Language}

\def\gpls{\glspl{GPL}\xspace}

\newacronym[longplural={Lines of Code}]{LoC}{LoC}{Line of Code}

\newacronym{EIS}{EIS}{Enterprise Information System}

\newacronym{PAIS}{PAIS}{Process-Aware Information System}

\newacronym{DTO}{DTO}{Data Transfer Object}

\newacronym{GUI}{GUI}{Graphical User Interface}

\newacronym{GUI DSL 1}{GUI DSL 1}{First version of GUI DSL}

\newacronym{GUI DSL 2}{GUI DSL 2}{Second version of GUI DSL}

\newacronym{CoCo}{CoCo}{Context Condition}

\newacronym{DT}{DT}{Digital Twin}

\newacronym{DS}{DS}{Digital Shadow}

\newacronym{CPPS}{CPPS}{Cyber-Physical Production System}

\newacronym{SLE}{SLE}{Software Language Engineering}
\def\sle{\gls{SLE}\xspace}

\newacronym{UI}{UI}{User Interface}

%% file: src/pre/00.abstract.tex
\begin{abstract}
		%
		In Software Language Engineering, there is a trend towards reusability by composing modular language components.
		However, this reusability is severely inhibited by a gap in integrating whitespace-sensitive and whitespace-insensitive languages.
		%
		There is currently no consistent procedure for seamlessly reusing such language components in both cases, such that libraries often cannot be reused, and whitespace-sensitive languages are developed from scratch.
		This paper presents a technique for using modular, whitespace-insensitive language modules to construct whitespace sensitive languages by pre-processing language artifacts before parsing.
		The approach is evaluated by reconstructing a simplified version of the programming language Python.		
		Our solution aims to increase the reusability of existing language components to reduce development time and increase the overall quality of software languages.
		
\end{abstract}

\keywords{Language Engineering, Software Language Composition, Language Reuse, Language Modularity, White-Space-Sensitivity}

%% file: src/tex/01.introduction.tex
\section{Introduction}
\label{sec:intro}


Software languages are a universal part of modern applications and systems.
They are employed in a wide variety of application domains, such as software development \cite{volter2013model}, automotive \cite{blom2013east}, avionics \cite{KPR+22}, or civil engineering \cite{visconti2021model}.
In these fields, they realize various aspects of a system, such as requirements, structure, and behavior.

Software languages can have different characteristics.
For example, they can be used for direct programming (C++ \cite{stroustrup1999overview}, Java \cite{arnold2005java}, Python \cite{van2007python}) or more abstract modeling (UML \cite{OMG17UML}, SysML (v2) \cite{OMG19SysML,OMG23SysMLv2}).
Additionally, there is a distinction between broadly applicable \gpls and \dsls tailored to specific application domains \cite{mernik2005DSLvGPL}. 
The overall discipline of constructing software languages is called \sle \cite{kleppe2008software}, where methods, techniques, and tools are investigated to manage the creation, maintenance, and evolution of these languages.

Since software languages are pieces of software in themselves, they face the same challenges in terms of ongoing development and maintenance \cite{favre2005languages}.
Therefore, \sle also has to address this challenge.
Deployment and reuse of libraries with modularly embeddable components was an important step in general software engineering \cite{frakes2005software}.
In recent years, this has also become increasingly important in \sle, where language composition techniques \cite{CCF+15b} and design patterns \cite{DJR22} for integrating existing languages have been developed.
Additionally, libraries of atomic language components also emerged \cite{BEH+20}, which serve as modular building blocks for literals, expressions, statements, etc.
By reusing such components, the development time can be drastically reduced while the quality of the product increases.

Over the years, various kinds of languages have been invented. 
A significant distinction in their creation stems from the design decision of whether a language should react sensitively to indentations.
Such a language is called whitespace-sensitive \cite{synytskyy2003robust}. 
This means that indentations and spaces directly influence the meaning of a program by indicating logical blocks.
Other languages that do not have this sensitivity regulate such logical coherences via special syntactic constructs, such as curly brackets.
The number of whitespaces in between has no significance here.

\begin{figure*}[ht]
	\centering
	\includegraphics[width=0.99\textwidth]{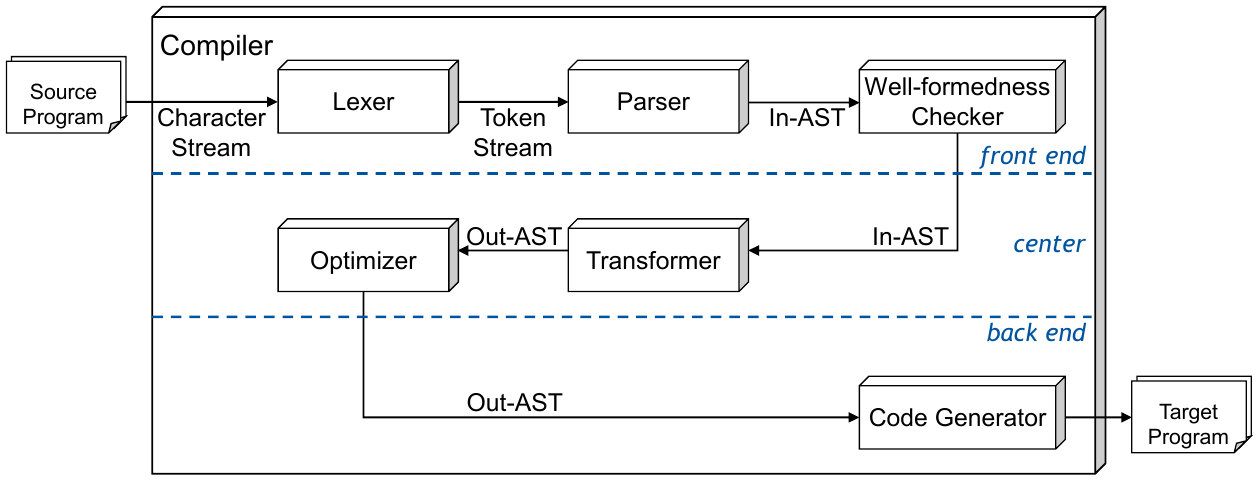}
	\caption{A general architectural overview of a compiler with its constituents.}
	\label{fig:general_compiler}
\end{figure*}

Due to their inherently different nature, whitespace sensitive and insensitive languages are usually developed completely separately from each other.
Thus, there is a research gap in the mutual embedding of such language components across this (in-)sensitivity barrier.
This impedes reusability in \sle, which ultimately leads to repeated, very similar implementations.
As a result, several existing language components cannot be used in whitespace-sensitive language developments.
Therefore, this paper addresses the research question: \emph{"How can whitespace-insensitive language components be used in the development of whitespace-sensitive languages?"}
We present a concept that eliminates the whitespace sensitivity of textual programs and models by means of preprocessing.
The preprocessed artifacts are then passed on to a parser composed of ordinal language modules for further processing. 
This bridges the gap between the two conceptual language worlds.
Furthermore, we demonstrate the applicability of our approach by developing a simplified version of the Python programming language, a prominent whitespace-sensitive example.

Thus, the contributions of our paper are: 

\begin{itemize}
  \item A concept for employing whitespace-insensitive language components in building whitespace-sensitive software languages 
  \item An Evaluation of the concept through the modular reconstruction of a version of the Python programming language
\end{itemize}

The remainder of this paper is structured as follows: 
\autoref{sec:preliminaries} introduces the necessary background knowledge on the examined topic and preliminary work for realizing the exemplary application.
In \autoref{sec:approach_general}, we present the general approach of employing preprocessors to bridge whitespace-(in-)sensitivity.
\autoref{sec:validation} showcases the development of a simplified version of Python and an evaluation of the approach's general applicability.
In \autoref{sec:discussion}, we discuss our solution.
\autoref{sec:related} considers related work, and finally, \autoref{sec:conc} concludes.

%% file: src/tex/02.background.tex
\section{Preliminaries}
\label{sec:preliminaries}

Our contribution is based on current research in \sle, compiler construction, and language composition. 
This section provides the fundamentals and introduces the language workbench MontiCore, in which technological space we conducted our case study and evaluation.

\subsection{Software Language Engineering}
\label{subsec:sle}

\sle is a general discipline focused on designing, developing, combining, and maintaining various kinds of software languages \cite{kleppe2008software}.
A software language defines a set of human-readable and machine-processable sentences.
Typically, they consist of a concrete and abstract syntax, additional well-formedness rules, and the semantics \cite{CGR09} of a language in terms of meaning \cite{HR04}.
The concrete syntax, i.e., the human-readable representation of a language, is usually textual or graphical.
A combination of both is also possible.
The abstract syntax is described by a data structure, such as a metamodel.
Defining textual languages has the advantage that a (often context-free) grammar (CFG) \cite{klint2005toward} in \ebnf \cite{wirth1996extended} allows concrete and abstract syntax to be defined together in one effort.
A graphical representation of the concrete syntax usually has to be defined additionally and mapped to the data structure of the abstract syntax.
Well-formedness rules express additional restrictions on the set of valid sentences with reference to a respective context.
The semantics of a language provide these valid sentences with a concrete meaning.
This requires a semantic domain into which the respective syntactic constructs must be mapped.
This mapping can be done both implicitly, through informal description, and explicitly, for example, via denotational semantics \cite{mosses1990denotational}.

In general, we differentiate between different types of languages.
\gpls can be used universally and across domains.
\dsls, on the other hand, are specifically tailored to an application domain and its concrete problem statements \cite{mernik2005DSLvGPL}.
When realizing the concrete syntax, the terminology of the corresponding discipline is often used.
In addition to this categorization, there is also a distinction between classic programming languages and modeling languages.
In general, both types can be both \gpls and \dsls.
It depends solely on their applicability and how they are employed in different application domains.
Therefore, these classifications are slightly orthogonal to each other.
Programs for programming languages are either evaluated in an execution engine or translated into direct machine code.
In recent years, more and more modeling languages have emerged that aim to abstract from the solution domain of software engineering and thus facilitate work in different use cases.
Models of these languages are often first translated into executable program code of a programming language.
\sle has brought up language workbenches \cite{erdweg2013state} such as Xtext \cite{eysholdt2010xtext}, MPS \cite{campagne2014mps}, and MontiCore \cite{HKR21}, which help develop such languages.
Based on an abstract language definition, such as CFGs or metamodels, such workbenches already generate a large part of the required infrastructure.

Such a processing structure of models or programs is very similar to the construction of classical compilers \cite{aho2008compiler}.
These process the (textual) input, convert it internally, and provide an output, either adhering to a programming language or in the form of direct machine code.
\autoref{fig:general_compiler} contains a schematic representation according to \cite{aho2008compiler} and \cite{cooper2022compiler_engineering}.
First, the input is passed to the lexer as a character stream.
In the following lexical analysis, predefined sequences of letters are translated into so-called tokens.
This assignment can be defined as regular expressions. 
The resulting token stream is passed to the parser, which translates the input into an AST using syntactic analysis.
There are different approaches to this transformation with different strengths and efficiencies, such as LL(k) \cite{parr1995antlr} or LR \cite{aho1974lr} parsing techniques.
All further well-formedness checks, analyses, transformations, and optimizations are performed on this tree representation of the program or model.
In some cases, an AST transformation translates the data structure closer to a representation suitable for synthesis.
It is often optimized for efficiency reasons and finally translated into the corresponding target language by a code generator.

\subsection{Language Composition}

Software languages become more and more sophisticated over the years.
However, this also brings an increasing demand for maintenance and evolution \cite{favre2005languages}.
Therefore, it has been explored how certain parts of existing languages can be reused \cite{Val10} such that new \sle efforts or when developing variants of existing languages do not always have to start from scratch.
This has resulted in various techniques for composing \dsls and \dsl fragments.
In this paper, and according to \cite{HLN+15a}, we distinguish between the following:

\paragraph{Language Inheritence}
The most basic variant is the inheritance of languages, whereby an existing language definition is incorporated.
Its productions are used, extended, or overwritten in the context of a new one.
This makes it easy to create several variants of a basic language without the need to modify it.
A special form of this composition technique is conservative extension.
Technically, it works like inheritance but with the methodological restriction of only adding optionally incorporable productions.
This means that models of the original language remain valid in the new language space.

\paragraph{Language Embedding}
An advanced form of composition is embedding, where several existing language definitions are combined into a new one. 
In this way, constructs of all incorporated languages are made accessible, and new ones are created due to their integration.
This is particularly powerful when language components are already designed for later embedding by providing common interfaces and explicit extension points.
A prominent example is the embedding of basic constructs such as expressions.
In the context of this work, language embedding is of particular relevance, as the combination of several language components requires their parsers to be integrated as well.
This directly impacts how models or programs of such integrated languages can be processed.

\paragraph{Language Aggregation}
The aggregation of languages is slightly different from the other composition techniques.
Here, the models or programs of the original languages remain separate artifacts, which can, however, reference each other.
Through symbolic linking and resolving the dependencies, they operate in a shared context.
This type of composition is, therefore, the loosest type of coupling and is particularly suitable for creating sophisticated language families from existing but mature languages \cite{HJK+23}.

\subsection{Whitespace-Sensitivity in Languages}

During the lexical analysis in a traditional compiler, the character stream of the input program or model is processed and tokenized (cf. \autoref{subsec:sle}).
There are two main approaches to dealing with whitespaces, tabs, or newlines. 
A widespread variant is that the lexer generally discards these characters so that they have no further syntactic or semantic influence on the program.
They only act as separators for other tokens and can be used for formatting to increase the readability of the source.
The contrary approach is to process these characters just like any other as part of tokens.
The appearance of these characters, therefore, leads to a different token sequence and, thus, to different results in the subsequent syntactical analysis.
Languages with this processing behavior are generally designated as whitespace-sensitive \cite{synytskyy2003robust}.
Here, indentations and newlines are used to actively shape the program flow.

\begin{figure}
  \centering
  \includegraphics[width=\columnwidth]{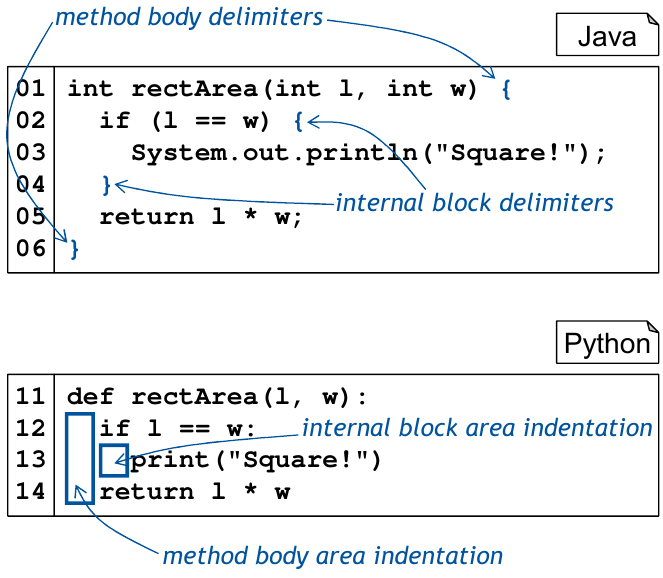}
  \caption{Comparison example of a program in the whitespace-insensitive language Java (top) and the whitespace-sensitive language Python (bottom).}
  \label{fig:jvpy}
\end{figure}

There are different advantages and disadvantages to using whitespace sensitivity in language design.
So far, there does not seem to be a definite favorite.
\autoref{fig:jvpy} contains a direct comparison between the two \gpls, Java and Python, the latter being whitespace-sensitive.
In both snippets, a function calculates a rectangle's area from a given length and width.
It also checks whether it is a square (ll. 02-04 and ll. 12-13).
In this case, a corresponding output is generated.
A side-by-side comparison shows that the Java version is somewhat more verbose than the Python version .
In Java, explicit delimiters (in the form of curly brackets) must be used to indicate the boundaries of the method body (ll. 01 and 06) or the internal block (ll. 02 and 04) caused by the if-statement\footnote{Please note that there is also a short notation for single-line if-statements in Java without an explicit delimiter. However, this has been omitted here for simplicity and is often considered a poor style in this language.}.
The Python variant is somewhat more compact. 
There are no delimiters here.
The content of the function body (ll. 12-14) and the internal section through the if condition (l. 13) are determined solely by the respective indentation.
Even if similar indentations are used in the Java case, this is pure syntactic sugar without any influence on the program flow.
Theoretically, the entire Java method could be written in a single program line, which is impossible in Python.

Although this is only a brief example without general significance, it can already be seen that whitespace-sensitive languages tend to be somewhat less verbose due to their nature.
This is caused simply by the fact that indentations are used in both cases for better readability, but the insensitive languages require additional delimiters.
One disadvantage of whitespace sensitivity, however, is an increased risk of errors due to slightly incorrect indentation. 
This is amplified by the fact that both normal spaces and tabs can generate indentations.
Python, for example, does not allow the mixing of both forms.
The scripting language Make \cite{feldman1979make} only allows tabs for indentation.
This can result in erroneous programs, even though they might appear syntactically correct.

\subsection{MontiCore Language Workbench and Reusable Language Component Library}

MontiCore is a language workbench for the efficient engineering and provision of textual DSLs \cite{HKR21}.
Based on a CFG, MontiCore automatically provides the necessary components of a model processing tool similar to a compiler (see Figure 1).
The language workbench generates lexers and parsers based on ANTLR technology \cite{parr1995antlr}.
Furthermore, an infrastructure for the implementation and execution of well-formedness rules, so-called context conditions, is created.
A customized visitor pattern \cite{DJR22}, generatively tailored to the language's abstract syntax, effectively enables the traversal of the AST for the analysis and execution of customized operations.
A symbol management infrastructure \cite{BMR22} also enables efficient navigation and cross-referencing within a model as well as beyond its artifact boundary.

MontiCore comes with a rich library of pre-built language components, ready for embedding \cite{BEH+20}.
These include, among others, different forms and levels of expressions, statements, literals, or types.
When designing the CFG, a language engineer can draw on these building blocks to efficiently incorporate concrete and abstract syntax, as well as their implemented tooling such as analyses, type checks, or generators.
Their suitability for use has already been demonstrated in various research and industrial projects \cite{DJR+19,HJRW20,DJRS22,KRS+22,HJK+23}.
However, all language components in the MontiCore ecosystem are whitespace-insensitive.
Their applicability for reuse in whitespace-sensitive cases is yet to be shown.




%% file: src/tex/03.approach.tex
\section{General Approach}
\label{sec:approach_general}


Many language component libraries are designed to efficiently support C-style languages, but they often lack native support for Python-style syntax. To make these components usable in the development of whitespace-sensitive languages, their frontend can be adapted to include a whitespace-insensitive parser, reducing the problem to the default case.
Indentation-based structures are handled by introducing a preprocessing step between the lexer and the parser, which replaces whitespace tokens with more meaningful ones that indicate indentation changes or statement boundaries.
With these new tokens, a language with significant whitespace can be described with a grammar that is similar to grammar defining a language without significant whitespace, with the difference that
blocks opened by delimiters, like the body of a method that is surrounded by curly parentheses, are instead delimited by the newly introduced control tokens.   

\autoref{lst:example_grammar} presents a grammar that imports a language component for expressions, which does not consider indentation, and defines tokens related to statements and blocks.
These tokens are later used to seperate multiple print and if statements during the parsing of the \texttt{Program} production.
Notably, the grammar omits token definitions for whitespace, resulting in a relatively compact description of the language.

To parse the program shown in \autoref{lst:example_program}, it must first be transformed by inserting \texttt{BLOCK\_START}, \texttt{BLOCK\_END}, and \texttt{STATEMENT\_END} tokens at appropriate locations, a preprocessing step described later in this section.
For illustrative purposes, \autoref{lst:example_program_processed} represents these tokens using familiar separators from C-style languages.
However, in practical applications, this representation is discouraged, as such tokens may already be in use elsewhere in the language, potentially leading to parsing ambiguities.

\begin{figure}
\begin{lstlisting}[label=lst:example_grammar,caption={Grammar for a whitespace sensitive language}]
import component Expressions;

token BLOCK_START, BLOCK_END, STMT_END;
token PRINT = "print";
// ...

Program = Statement*;
Statement = PrintStatement | IfStatement;
PrintStatement = PRINT STRING STMT_END;
IfStatement = IF Expression COLON BLOCK_START Statement+ BLOCK_END;
\end{lstlisting}
\end{figure}

\begin{figure}
\begin{lstlisting}[label=lst:example_program,caption={A simple program in that language}]
print "Hello"
if 1 < 2:
	print " world"
\end{lstlisting}
\end{figure}

\begin{figure}
\begin{lstlisting}[label=lst:example_program_processed,caption={The preprocessed version of that program}]
print "Hello";
if 1 < 2:{
	print " world";
}
\end{lstlisting}
\end{figure}


\subsection{Language frontend}
\label{subsec:language-frontend}
\begin{figure*}
	\includegraphics[width=\linewidth]{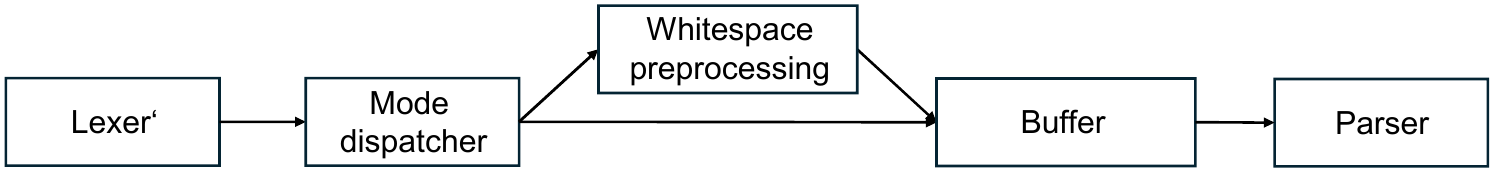}
	\caption{Components of the transformed language frontend}
	\label{fig:frontend-transformation}
\end{figure*}

Typically, lexers for whitespace-insensitive languages discard all whitespace tokens to drastically simplify the grammar describing the parser.
Thus, the lexer of the original frontend cannot be reused as is.
It needs to be adapted to also emit relevant whitespace tokens,
such that the \texttt{Whitespace preprocessing} component, introduced in in \autoref{fig:frontend-transformation}, can process them and adapt the token stream.
In some parser generators, this is a small configuration change, like in Antlr \cite{parr1995antlr}, where this can achieved by sending all whitespace to a specific channel of the token stream.
In others, it might be a larger obstacle to using this process. 

\begin{figure}
\begin{lstlisting}[label=lst:example_indent_irrelevant,caption={Method call parenthesis in Python make indents irrelevant}]
print(
	"Hello",
		"World",
"!"
)
\end{lstlisting}
\end{figure}

As depicted in \autoref{lst:example_indent_irrelevant}, some whitespace-sensitive languages also contains productions where the indentation is not relevant.
These are typically delimited by easy to identify tokens, like opening and closing parenthesis. 
In that case, an additional component \texttt{ModeDispatcher} can be added to the frontend pipeline, which dispatches all tokens either to the \texttt{Whitespace preprocessing} previously described, or passes them to the parser unchanged. 

To reuse the parser without adaption, the input type needs to remain a single stream of tokens.
Thus, tokens resulting from both the indentation-sensitive and -insensitive processing are combined in the \texttt{Buffer} before being passed to the parser.
The buffer also hides from the parser the fact that a single raw token from the lexer can result in no token (in the case of removed whitespace), a single original token, or multiple tokens, i.e., the original token followed by a control token.

While preprocessing before lexing has been proposed as an alternative, e.g., by Fowler \cite{Fowler10},
this approach has some drawbacks when compared to processing the whitespace after lexing.
A naïve preprocessing step without a lexer cannot accurately account for whitespace occurring within other tokens, such as multiline strings.
These cases would require a second lexer implementing at least a subset of the language's lexical rules.
Consequently, the most pragmatic trade-off appears to be adapting the original lexer to emit relevant whitespace tokens.
Nevertheless, in scenarios where processing the indentation before the lexer is desirable, such as in languages without multiline strings, the described pipeline remains applicable with some changes.
The preprocessed token stream can be serialized back into a character stream and subsequently fed into the original, unmodified lexer.

\subsection{Whitespace preprocessing}
\autoref{lst:preprocessing} contains the base algorithm to convert indentation into appropriate control tokens.
To add them at the correct positions, the preprocessor needs to keep track of the current indent,
as well as the number of blocks currently opened.
If it encounters a line break, it counts the indent of the next line by looking ahead in the token stream
until a non-indent token is found.
This lookahead only creates a small overhead, as it is only performed once per encountered newline token and does not result in additional backtracking in the parser.
If the indent stays the same or decreases, a \texttt{STMT\_END} token is emitted, which signals to the parser that the current statement is finished.
Based on the increase or decrease in indent, a corresponding \texttt{BLOCK\_START} or \texttt{BLOCK\_END} token is also emitted.
If the end of file is observed, all currently open statements and blocks need to be closed by emitting corresponding control tokens.

Note, that this algorithm needs to be adapted if the language contains blocks that cannot contain statements.
In that case, the decrease of indent or end of file may not correspond to the end of a statement. 

Another special case is line continuation, which allows a line of the program as seen by the parser to be split into multiple lines in the actual text of the document.
In this algorithm, it is handled by ignoring the \texttt{continue\_line} token, which must include the line break.
Otherwise, additional state to ignore the next newline token needs to be introduced.

\begin{lstlisting}[mathescape,label=lst:preprocessing,caption={Injection of tokens},float]
oldIndent = 0
blockDepth = 0

method handleToken(t, stream):
	if t == $t_{continue\_line}$:
		pass

	else:
		if t $\in T_{newline}$:
			newIndent = calcCurrentIndent(stream)

			if newIndent > oldIndent:
				blockDepth++
				emit($t_{block\_start}$)
			else if newIndent < oldIndent:
				blockDepth--
				emit($t_{stmt\_end}$)
				emit($t_{block\_end}$)
			else:
				emit($t_{stmt\_end}$)

			oldIndent = newIndent

		if t == $t_{EOF}$:
			emit($t_{stmt\_end}$)
			for i in 0..blockDepth:
				emit($t_{block\_end}$)

		emit(t)
\end{lstlisting}

\subsection{Mode Dispatcher}
\begin{figure*}
	\centering
	\hspace{0.2\textwidth}
	\includegraphics[width=0.8\textwidth]{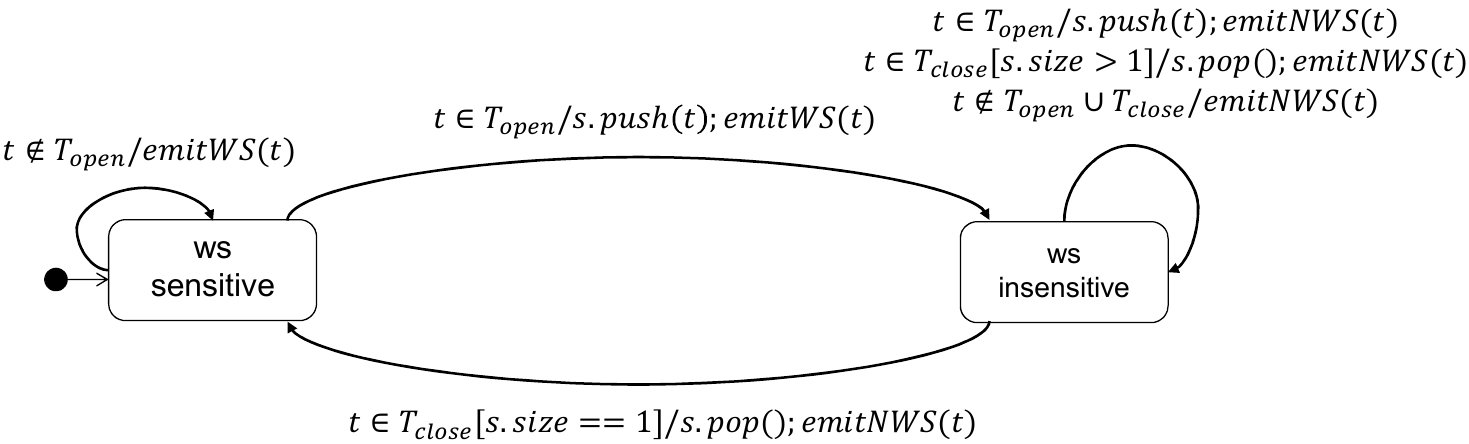}
	\medskip
	\medskip
	\caption{Behaviour of the Mode dispatcher stage of the frontend}
	\label{fig:mode_dispatcher_statechart}
\end{figure*}

The behaviour of the \texttt{ModeDispatcher} component is described as an statechart in \autoref{fig:mode_dispatcher_statechart}.
Either the pipeline currently needs to consider whitespace, then it is in the state \texttt{ws sensitive}, or it does not, in state \texttt{ws insensitive}.

Any token $t$ that does not open a indentation neutral block($t \notin T_{open}$), will be passed along to the \texttt{Whitespace preprocessing} using the method \texttt{emitWS}, without moving to the \texttt{ws insensitve} state.
Otherwise, the state is switched and the token is pushed to the stack $s$.
This addition is necessary to handle nesting of sections that are whitespace-insensitive.

While in the \texttt{ws insensitive} state, the component continues to forward all tokens directly to the buffer using the \texttt{emitNWS} function. If an additional block-opening token $t \in T_{open}$ is encountered, it is pushed onto the stack $s$, allowing proper handling of nested indentation-neutral blocks, and the token is emitted.

If a block-closing token $t \in T_{close}$ is received while the stack size is greater than 1 ($s.\text{size} > 1$), the top of the stack is popped and the token is emitted, but the state does not change. This ensures that only the outermost insensitive block controls the transition back to the \texttt{ws sensitive} state.

Once the final closing token is encountered, i.e., $t \in T_{close}$ and $s.\text{size} == 1$, the stack is popped, the token is emitted, and the state is switched back to \texttt{ws sensitive}. Tokens that are neither in $T_{open}$ nor in $T_{close}$ are simply emitted in-place to the corresponding component of the pipeline in both states without causing any transitions.

This design allows the component to correctly control the sensitivity to whitespace depending on the current syntactic context, while also supporting arbitrarily nested insensitive regions.
Note, that matching of pairs of opening and closing tokens is left to the parser, since theses rules should already be described by the productions of the language. 

\subsection{Requirements to Language Components}
The presented approach aims to reuse whitespace-insensitive language components to build whitespace-sensitive languages.
To achieve this, the components need to fullfil the requirements described in the next portion of the paper.

\paragraph{RQ1}
\label{req:rq1}
When a component relies on significant whitespace, it must not process whitespace in the parser but define control tokens as decribed above, since the proposed frontend discards all whitespace before it reaches the parser.
Additionaly, extension points to configure these control tokens need to be available.
Otherwise, when multiple such components are used in conjunction, two or more diffent tokens could have the same meaning, e.g., an increase in indent.
This forces the preprocessor to track which parts of the constructed language require which tokens, increasing complexity and undermining modular reuse.
Harmonizing control tokens via extension points allows the preprocessing step to remain simpler and more robust.

\paragraph{RQ2}
For components that include whitespace-insensitive regions, these areas should be clearly delimited in a way that is easy to identify in the token stream, e.g., through counting opening and closing parentheses.
Otherwise, the conversion into the insensitive format can introduce a hefty performance penalty.
If the rules to delimit such regions are complex enough to require backtracking, then the presented approach might not be usable at all, and the whitespace processing must be shifted into the parser.
Backtracking often involves revisiting and reinterpreting earlier characters, which conflicts with the assumption of a fixed, immutable token stream typically required by parsers.

\vspace{\baselineskip} 
Many language components, such as literal and expression definitions, can be reused directly, as they typically do not introduce indentation-sensitive structure (cf. RQ1) and either do not span whitespace-insensitive regions at all or delimit them in a straightforward manner (cf. RQ2).
For example, string and numeric literals are inherently insensitive to whitespace and occur in isolation, while collection literals (e.g., arrays, sets, maps) often use syntactic markers like brackets that clearly bound the region.
This same reasoning applies to most expressions, with the exception of constructs like lambda expressions, which may introduce nested, indentation-sensitive bodies.
In contrast, components that define hierarchical structure, such as control flow constructs, type declarations, or method definitions, require adaptation by adding extension points to meet RQ1.

%% file: src/tex/04.evaluation.tex
\section{Validation}
\label{sec:validation}


To verify that the proposed approach works, a parser for a subset of the Python language is implemented using MontiCores library of language components.
These components are not natively whitespace-sensitive, since they target the creation of C-style languages.

To identifiy relevant parts of the library, the official Python grammar\cite{van2014python} is compared to the available components.
The expression components, \texttt{Common\-Expressions} and \texttt{Assignment\-Expressions}, as well as a components defining literals and basic identifiers, \texttt{MCCommonLiterals} and \texttt{MCBasics}, contain productions that can be reused and thus are included.
Non of the components defining statements can be reused, since they are constructed similar to Java, and thus,
include semicolons at the end of statements, rounded parentheses around conditions, and curly parentheses around blocks.
Even though MontiCore supports overriding tokens when using language components, potentially allowing replacement of undesirable ones, it does not allow overriding them only in certain productions.
Thus, the decision was made to rewrite all statements for this implemtation, to get the necessary level of control.    



As the first step of defining the grammar, the control tokens are chosen.
To enable easier inspection of the token stream, printable unicode characters that are not valid in Python outside of strings are used to avoid overlap with existing tokens.
The left white curly bracket(U+2983) and right white curly bracket(U+2984) - looking similar to curly braces - are used to indicate a block start and end respectively.
To signal the end of a statement, the reversed semicolon(U+204F) is used.
Since Python itself allows statements to be seperated by a semicolon, it is added as a second alternative to this token definition. 

Next, additional literals and expressions that are not supplied by the components are implemented in new token rules and productions. 
The basic structure of a Python artifact is implemented as a list of statements, using MontiCores interface production mechanism to define the production \texttt{Statement}.
Standard statements, like the \texttt{if} and \texttt{while}, as well as Python specific statements, like \texttt{with}, are added as implementations of this interface production and incorperate the control tokens where appropriate.

Since no used component uses control tokens for whitespace processing, the language component \texttt{BasicStatements}, displayed in \autoref{lst:basic-statements}, is extracted from the main grammar.
It contains the standard statements that are not specific to Python.
As prescribed by requirement RQ1 of \autoref{req:rq1}, it defines the extension points \texttt{StartBlock}, \texttt{EndBlock}, and \texttt{EndStmt} to be filled by the appropriate tokens in the importing grammar.

\begin{figure}
\begin{lstlisting}[label=lst:basic-statements,caption={Excerpt from the extracted \texttt{BasicStatements} language component}]
component grammar BasicStatements extends MCBasics, ExpressionsBasis {
 external StartBlock;
 external EndBlock;
 external EndStmt;

 interface Statement;

 scope StatementBlock = (StartBlock StatementBlockBody EndBlock) | Statement;
 StatementBlockBody = Statement+;

 IfStatement implements Statement = "if"     condition:Expression     ":" thenStatement:StatementBlock
          ("elif" elifCondition:Expression ":" elifStatement:StatementBlock )*
          ElseStatementPart?;

 ElseStatementPart = "else" ":" StatementBlock;

 // ...
}
\end{lstlisting}
\end{figure}

From the components and the grammar, MontiCore generates a complete, ANTLR-based frontend for the whitespace-insensitive version of the language, including a parser and lexer.
To implement the necessary preprocessing to derive a parser of the whitespace-sensitive language, a subclass of ANTLRs \texttt{Token\-Source} is implemented and passed to the parser instead of the original lexer. 
As described in \autoref{subsec:language-frontend}, the next step is to adapt the lexer to emit whitespace tokens, which in this case is a trival change, since lexers generated using MontiCore \texttt{MCBasics} component include a second channel in the token stream dedicated to whitespace.

The description of Pythons lexer\cite{van2014python} states that expressions can be split across multiple lines, without using the line continuation marker("\textbackslash"), as long as they are enclosed by round, curly, or square parentheses.
In these cases, the indention becomes irrelevant until all parenthesis pairs are closed.
Thus, we implement the \texttt{ModeDispatcher} component as described in \autoref{fig:mode_dispatcher_statechart} and define $T_{open}$ as the tokens for \texttt{'('}, \texttt{'['}, and \texttt{'\{'}.
$T_{close}$ is equal to the tokens for \texttt{')'}, \texttt{']'}, and \texttt{'\}'}.
All tokens produced by \texttt{emitWS} are processed using the class \texttt{Whitespace\-Sensitive\-Processor},
while all others are directly passed to the buffer. 

Similarly, the \texttt{Whitespace preprocessing} algorithm from \autoref{lst:preprocessing} is implemented in the class \texttt{Whitespace\-Sensitive\-Processor}.
To avoid overlap in token defintions, the line continuation character("\textbackslash") is defined without capturing trailing whitespace.
Thus, the fact that a line is continued needs to be kept in the state of the preprocessor
and is used to discard the next newline token once it arrives.

The \texttt{Buffer} component is implemented in the class \texttt{PreprocessingTokenSource} and
combines the tokens that are either unprocessed or processed based on the current state of the \texttt{ModeDispatcher}.
Additionaly, it hides all whitespace tokens from the parser by moving them to a seperate channel of the token stream.
This is necessitated by the configuration changes made to the lexer.

Now that all components from section \autoref{sec:approach_general} are implemented
and integrated into the pipeline, a whitespace-sensitive frontend for a version of Python has been created using whitespace-insensitive language components.
Even though some less used language constructs are left out of the grammar for brevity, the parser can successfuly parse most files of bigger python projects, demonstrating that the concept works.
GemPy\cite{de2019gempy} consists of 111 Python files, all of which can be parsed without errors.
In contrast, 102 of the 3367 Python files in the Transformers library\cite{wolf2020transformers} are parsed with errors, most of which are related to missing concepts, not whitespace processing.


\subsection{Limitations of the resulting frontend}
Some expressions that are provided in the used language components are not valid in python, e.g., the \texttt{Inc\-Prefix\-Expression}.
This expression parses \texttt{++i}, which increments the value of the variable \texttt{i} and returns the new value.
A possible solution is to override the offending productions in the grammar to throw an error if it is successfuly parsed, e.g., in MontiCore syntax: \lstinline|IncSuffixExpression implements Expression = Expression "++" {throw new RuntimeException("IncSuffixExpression not allowed");};|. 
Alternatively, the resulting \ast can be inspected after parsing and rejected before further processing if it contains invalid expression types.
This is the approach chosen here and easily implemented using the generated context condition infrastructure generated by MontiCore.

%% file: src/tex/06.discussion.tex
\section{Discussion}
\label{sec:discussion}

We have demonstrated the reuse of whitespace-incentive language components in the compositional language construction of whitespace-sensitive languages.
Our approach is based on preprocessing, which is done by invasively modifying the lexical analysis within a compiler.
The presented algorithm is technology-independent and represents a constructive solution to the research question.
The exemplary application of the technique in the MontiCore ecosystem makes the existing language components available for an extended portfolio of language definitions.
However, the reverse direction, i.e., the use of whitespace-sensitive language components in whitespace-insensitive languages, is still unsolved.
A reverse preprocessing approach to the one presented here, in which indentations are recognized and replaced by corresponding language-specific delimiter tokens, is conceivable.
The exact procedure, as well as proof of viability, still needs to be provided.
Arguably, however, we could claim that the most relevant direction has been elaborated in this paper since, to the best of our knowledge, no reusable library of indentation-sensitive language components currently exists.

Although our approach is a general solution, a few threats to validity remain in terms of generalizability \cite{WRH+12}.
A preprocessor, as described, must be rebuilt for each composition scenario and cannot be reused in a backbox fashion.
Unfortunately, the presented approach is dependent on the underlying languages and their components in two aspects.
On the one hand, the incorporated language components specify exactly which characters serve as delimiters.
Therefore, the preprocessor must adaptively add precisely these correct delimiter symbols to the character stream.
This problem can be considered minor, as this only has to be done once for the set of included components.
Furthermore, the variance between these special delimiter characters is most likely relatively small since modern languages (cf. Java \cite{arnold2005java}, C \cite{hejlsberg2008c}, C++ \cite{stroustrup1999overview}, C\# \cite{hejlsberg2006csharp}, JavaScript \cite{koch2011javascript}) often use similar characters for this purpose.
The second problem is the dependency on the underlying language to be parsed.
There are different ways of indenting, for example via spaces or tabs.
Which characters may be used at all and how these are translated into delimiters in particular cases is language-specific and must, therefore, be implemented anew for the preprocessor each time.

Our approach can be used in any language workbench based on parser technology.
In our evaluation, we also use a parser based on ANTLR, which additionally facilitates the applicability in other technological spaces due to its wide distribution.
The solution presented here is mainly based on an intervention in the lexical analysis.
For use in scannerless parsing \cite{visser1997scannerless}, where the lexical analysis is minimized, our approach would have to be further elaborated.
Shifting the task to the parser makes the adaptive preprocessing step much more difficult, as black-box parser reuse is not possible anymore. 
A further obstacle to applicability is using purely projective editors, as provided by the MPS framework \cite{campagne2014mps}.
Without parsing technology at all, our approach cannot be applied.
However, since languages and language composition are generally realized differently in projectional editing \cite{voelter2012language}, new concepts must be developed for these cases.

Finally, we have shown how purely whitespace-insensitive language components can be used to construct whitespace-sensitive languages.
Whether these languages themselves can be composed well and thus employed in even larger embedding contexts is still unknown.



%% file: src/tex/05.relatedWork.tex
\section{Related Work}
\label{sec:related}

While our approach specifically aims at improving compositional language construction across whitespace sensitivity, a few related approaches exist for this or similar challenges.

Fowler proposes a similar attempt using preprocessing of the input program \cite{Fowler10}.
The main difference lies in the point in time of the preprocessing step.
In our approach, preprocessing takes place after lexical analysis, which implies a modification of the token stream.
On the other hand, Fowler describes a modification of the character stream i.e., before the lexical analysis.
Both approaches have different advantages and disadvantages.
Our solution has the advantage that it can handle whitespaces in tokens (such as multiline strings) without additional effort.
A modification at the character stream level inevitably assumes that no spaces are allowed in tokens.
Since this is generally not the case, this entails a major adaptation of the lexer.
However, Fowler's solution has the advantage of better supporting scannerless parsing concepts since our approach to modifying the token stream relies on prior lexical analysis.

Another approach applies a similar method for robust parsing of multilingual technologies on the web \cite{synytskyy2003robust}.
The method presented here involves a combination of preprocessing and island grammars \cite{moonen2001islandgr}. 
However, the goal is rather different.
There, island grammars are used to parse and analyze all incorporated programmatic parts in web applications, also taking into account potential syntactic errors.
A real composition in the sense of language embedding is neither targeted nor achieved within this work.

Another work on modular grammar specification \cite{johnstone2014modular} follows the same goal as our work of composing multiple language definitions in the form of grammars.
This approach is based on importing language modules by copying and integrating the corresponding nonterminal productions.
Each module can have its own indentation rules.
A corresponding algorithm is proposed to switch between these indentation rules for composing the modules into a combined grammar.
This approach promises an adaptive adjustment to the corresponding whitespace-(in-)sensitivity.
Due to the copying approach, however, the resulting grammar can quickly become very verbose, which can limit its readability and further composability.
Furthermore, complete integration includes not only the composition of the concrete syntax structures but also that of the generated and handwritten tooling.
Whether the approach presented in this paper can take integrating tooling into account is questionable.
Since our approach does not copy language definitions but integrates them at both syntax and implementation levels, we can reuse existing tooling.

In \cite{amorim2018declarative} a technique to describe layout-sensitive languages is presented,
which abstracts away from low level details of the whitespace processing.
If it is integrated into language workbenches and specific language components,
this could be used to harmonize the concrete whitespace handling of the combined language.
Since this approach works by adding annotations to productions of the grammar,
it can not be applied to reuse existing, whitespace-insensitive components to build whitespace-sensitive languages without manual adaption.
In contrast, with our approach some existing components fitting these criteria can be reused without modification.

%% file: src/tex/07.conclusion.tex
\section{Conclusion}
\label{sec:conc}

In this paper, we demonstrated a constructive approach to utilize existing whitespace-insensitive language components in the compositional language construction of indentation-sensitive languages.
Our solution is based on preprocessing the character stream in the lexical analysis of a general compiler.
By adaptively recognizing indentations in the source code and replacing them with corresponding delimiters from the embedded language components, such languages can be effectively realized.
We have evaluated this by rebuilding a simplified version of the Python programming language from modular components of the MontiCore language library.
The presented algorithm can generally be applied to compiler preprocessors. Nevertheless, there is still potential for future research in the area of scannerless parsing and application to projectional languages.
In general, our solution contributes to bridging the technological gap between whitespace-sensitive and whitespace-insensitive software languages and represents a further step in reusability within the SLE research field.